\documentclass[aps,prb,preprint,floats]{revtex4}
\usepackage{epsfig,amssymb,amsfonts,amsmath}
\usepackage{graphicx}

\input{epsf}
\input{tcilatex}

\begin{document}

\title{On exotic sphere fibrations, topological phases, and edge states in
physical systems}
\author{Hai Lin and Shing-Tung Yau}
\affiliation{Department of Physics, Harvard University, MA 02138, USA\\
Department of Mathematics, Harvard University, MA 02138, USA}

\begin{abstract}
\vspace{1cm}

We suggest that exotic sphere fibrations can be mapped to band
topologies in condensed matter systems. These fibrations can
correspond to geometric phases of two double bands or state vector
bases with second Chern numbers $m+n$ and $-n$ respectively. They
can be related to topological insulators, magneto-electric
effects, and photonic crystals with special edge states. We also
consider time-reversal symmetry breaking perturbations of
topological insulator, and heterostructures of topological
insulators with normal insulators and with superconductors. We
consider periodic TI/NI/TI/NI$^{\prime}$ heterostuctures, and
periodic TI/SC/TI/SC$^{\prime}$ heterostuctures. They also give
rise to models of Weyl semimetals which have thermal and
electrical transports.

\end{abstract}

\maketitle

\vspace{2cm}

\vspace{2cm}

\vspace{2cm}

\vspace{2cm}

\vspace{2cm}

\vspace{1cm}

\vspace{2cm}




\section{Introduction}

\label{sec_ introduction} In this article, we propose to realize the exotic
spheres as the geometric phases in condensed matter systems. We suggest that
the exotic sphere fibration can be realized as the geometric phases in
condensed matter systems, cold atomic or molecular systems.

The topological description of the quantum states of matter gives
a new method in describing the condensed matter. Condensed matter
systems with band structures that have nontrivial topological
properties give a new type of materials with properties that are
robust under certain perturbations. There are many interesting
topological properties for example the appearance of edge states
and the existence of gapless surface states.

The topological insulator has a bulk gap, while it has topologically
protected edge states on the boundary of the topological insulator. Thus,
the topological insulator is an insulator in the bulk while has gapless edge
states on the boundary of the topological insulator. The 3D topological
insulators have been shown in materials for example,\cite{Xia etal,Zhang
etal,Chen Y L etal,Hsieh et al} \textrm{Bi}$_{2}\mathrm{Te}_{3},~$\textrm{Bi}%
$_{2}\mathrm{Se}_{3},$ \textrm{Sb}$_{2}\mathrm{Te}_{3},~$\textrm{Bi}$_{1-x}%
\mathrm{Sb}_{x}$.$~$The 2D topological insulators have been observed in
\textrm{HgTe} quantum wells.\cite{Konig et al, Bernevig et al}

For the 3D topological insulators, the topologically protected
surface state realizes itself by the non-trivial spin texture on
the surface band of the topological insulator. The topological
surface state is protected by time-reversal symmetry. The
existence of odd number of or single surface Dirac point is robust
in the presence of nonmagnetic impurities, and other time-reversal
symmetry preserving perturbations. The difference between
topological insulator and normal insulator can be distinguished by a $%
\mathrm{Z}_{2}$ invariant.\cite{Fu et al, Moore Balents, Roy} The
3D topological insulators can be described\cite{Qi et al} by
topological field theory with a $\theta $ variable. Due to the time reversal symmetry, $%
\theta $ takes values of 0 or $\pi ,$ modulo $2\pi $. The $\theta $ term
gives magnetoelectric effects, with magnetoelectric polarization in the
materials.\cite{Qi et al,Essin et al} The spin polarization of surface band
and magnetoelectric polarization of the materials, can be experimentally
measured.

In early days of differential topology, John Milnor constructed a seven
dimensional compact space called exotic sphere. The space he constructed has
the property that it has a continuous one to one map to the round sphere,
and yet it cannot be mapped to the round sphere smoothly. There are 28
Milnor exotic spheres that are mutually distinct from each other. The
construction was based on distinct bundles over the four dimensional
manifold. Bundles over four dimensional manifold may be used to describe
geometric phases in condensed matter and atomic systems. It is interesting
to see whether one can build the Milnor exotic spheres into the theory of
condensed matter and atomic systems. It is also nice to connect subjects in
mathematics to subjects in theoretical physics.

In this paper, we suggest that exotic sphere fibrations can be
mapped to band topologies in condensed matter systems. These
fibrations can be mapped to geometric phases of two double bands
or state vector bases with their associated second Chern numbers.
They can be related to topological insulators, magneto-electric
effects, and photonic crystals with special edge states, among
other aspects. It is nice to understand the physics of topological
insulator in the situations when it is placed adjacent to other
kinds of materials. We also consider time-reversal symmetry
breaking perturbations of topological insulator, and
heterostructures of topological insulators with normal insulators
and with superconductors.

The organization of this article is as follows. In section \ref{sec_
geometric phase band chern number}, we discuss that exotic sphere fibrations
can be related to band topologies in condensed matter systems. These
fibrations can be mapped to geometric phases of two double bands or state
vector bases with the second Chern numbers $m+n$ and $-n$ respectively. In
section \ref{sec_ topological insulators}, we discuss their relation to
topological insulators, and magneto-electric effects. In section \ref{sec_
photonic crystals}, we also discuss their relation to photonic crystals with
special edge states. In section \ref{sec_ heterostructure TI/NI/TI/NI'}, we
consider time-reversal symmetry breaking perturbations of topological
insulator, and heterostructures of topological insulators with normal
insulators, and periodic TI/NI/TI/NI$^{\prime }$ heterostuctures. In section %
\ref{sec_ heterostructure TI/SC/TI/SC'}, we consider heterostructures of
topological insulators with superconductors, and periodic TI/SC/TI/SC$%
^{\prime }$ heterostuctures. These structures also give rise to models of
Weyl semimetals which have thermal and electrical transports. In section \ref%
{sec_ other relevant materials}, we also suggest relevance to other possible
materials such as cold atom systems and semiconductor systems.

\section{Geometric phases of two double bands or state vector bases and
related topologies}

\label{sec_ geometric phase band chern number} One of the
interesting types of fiber bundles are 3-sphere bundles over
4-sphere. Such fibrations can be constructed by patching two
$R^{4}\times S^{3}$ and identify their overlapping region by a
diffeomorphism. One can divide $S^{4}$ into three regions: a north
patch $R_{(1)}^{4};~$a middle patch $[-\epsilon ,\epsilon ]\times
S^{3};~$and a south patch $R_{(2)}^{4}.$ The $R_{(1)}^{4}\times
S^{3} $ can be parametrized by a quaternion $u$ and a unit norm
quaternion $v$, in which the $u~$belongs to the $R_{(1)}^{4}$ and
the $v~$belongs to the $S^{3}$. Similarly, the $R_{(2)}^{4}\times
S^{3}$ can be parametrized by a quaternion $u^{\prime }$ and a
unit norm quaternion $v^{\prime }$.$~$The transition function is
defined on the middle patch, and it is
\begin{equation}
u^{\prime }=\frac{u}{\left\Vert {\small u}\right\Vert ^{2}}%
,~~~~~~~~v^{\prime }=\frac{u^{m}(u^{n}vu^{-n})}{\left\Vert {\small u}%
\right\Vert ^{m}}=\frac{u^{n+m}(v)u^{-n}}{\left\Vert {\small u}\right\Vert
^{m}}.
\end{equation}%
$\left\Vert {\small u}\right\Vert $ denotes the norm of the quaternion $u$,
while \vspace{1pt}$\left\Vert v^{\prime }\right\Vert $ =$\left\Vert
v\right\Vert =1.$ Such fibrations can be classified by two integers $%
(n+m,-n).$

The fibration of $S^{3}~$over $S^{4}~$can be characterized by the map from
the middle patch $S^{3}$ to the structure group $SO(4)$ which corresponds to
the rotational symmetry of the $S^{3}~$fiber. This map is characterized by
homotopy group $\pi _{3}(SO(4))\cong \mathrm{Z}\oplus \mathrm{Z,~}$where $%
\pi _{3}(SU(2))\cong \mathrm{Z}$, and $so(4)=su(2)_{(1)}\times su(2)_{(2)}$.
This fibration is characterized by two integers, which correspond to $%
(n+m,-n).~$The two integers $n+m~$and $-n~$correspond to the second Chern
numbers of the $su(2)_{(1)}$ and $su(2)_{(2)}$. This may be viewed as%
\begin{eqnarray}
c_{2}^{(1)} &=&\frac{1}{8\pi ^{2}}\int ~(\mathrm{tr}F^{(1)}\wedge
F^{(1)})=n+m,  \label{c_1} \\
c_{2}^{(2)} &=&\frac{1}{8\pi ^{2}}\int ~(\mathrm{tr}F^{(2)}\wedge
F^{(2)})=-n,  \label{c_2}
\end{eqnarray}%
and the integration is on the four dimensional base manifold.

This can be interpreted as $n+m~$instantons of the $su(2)_{(1)}~$gauge~fields%
$,$ and $n$ anti-instantons of the $su(2)_{(2)}$ gauge fields.$~$This can
also be viewed as $m$ instantons of $su(2)_{(1)},$ plus $n$ pairs of $%
su(2)_{(1)}~$instanton and $su(2)_{(2)}~$anti-instanton. For
$m=1$, different $n$ are in different diffeomorphism classes, but
in the same homeomorphism class. The standard sphere corresponds
to $n=0$,$~m=1$, which
also corresponds to one instanton of $su(2)_{(1)}~$on $S^{4}$.$~$For $n>0$,$%
~m=1,~$it is the exotic sphere, and it also corresponds to $n+1$ instantons
of $su(2)_{(1)},~$and $n$ anti-instantons of $su(2)_{(2)}.~$The case for $%
m=2,$ and other general $m,$ are also very interesting.

We propose to realize exotic sphere fibrations (\ref{c_1}, \ref{c_2}) by
geometric phases in condensed matter systems, cold atomic systems or
molecular systems. These exotic fibrations can be mapped to band structures
of those systems with nontrivial band topologies in the systems. \

We start with wavefunction $\left\vert \psi ^{I}\right\rangle ,$ where $I$
labels bands in band structure, or labels state vector bases, and define
gauge field associated with the geometric phase in parameter space,
\begin{equation}
A_{a}^{(1)ij}=-i\left\langle \psi ^{i}(\{\xi _{a}\})\right\vert \partial
_{\xi _{a}}\left\vert \psi ^{j}(\{\xi _{a}\})\right\rangle ,
\end{equation}%
\begin{equation}
A_{a}^{(2)\alpha \beta }=-i\left\langle \psi ^{\alpha }(\{\xi
_{a}\})\right\vert \partial _{\xi _{a}}\left\vert \psi ^{\beta }(\{\xi
_{a}\})\right\rangle ,
\end{equation}%
where $i,j,$ $\alpha ,\beta ~$label different bands, or state vector bases. $%
\xi _{a}$ are parameters for the wavefunctions. The $i,j$ labels a double
band and takes values 1 or 2. The $\alpha ,\beta ~$labels a different double
band and takes values 1 or 2. We have written the states in orthonormal
basis. When the two bands are degenerate or almost degenerate, the geometric
phase becomes a non-abelian matrix-valued quantity. $\xi _{a}~$are the
coordinates on the parameter space. $(\xi _{1},\xi _{2},\xi _{3},\xi _{4})~$%
parametrize four dimensional manifold. Since there are two $su(2)$ gauge
fields, the band structure is such that there are four occupied bands, and
there are two $su(2)$ gauge fields associated with the two double bands in
the parameter space.$~$The two bands in the double band will be degenerate
at some points of the parameter space, the~$\xi $ space.

The four-parameter space can be momentum space, or it can be momentum space
together with extra parameters, or it can be other parameters in the model.
It is good to embed a three dimensional experimental system into a four
dimensional parameter space. For example, the four dimensional parameter
space $(\xi _{1},\xi _{2},\xi _{3},\xi _{4})\ $can be three dimensional
momentum space $(k_{1},k_{2},k_{3})$ with an additional parameter $\xi
_{4}.~\xi _{4}$ can be a parameter in the model Hamiltonian or the effective
model of the experimental system. $\xi _{4}$ can also be $k_{4},~$or $\omega
,~$or~a parameter in the model Hamiltonian. It can also be two dimensional
momentum space $(k_{1},k_{2})$ with additional parameters $\xi _{4},\xi
_{3}.~$It can also be real space, together with additional parameters.

The geometric phase is sometimes also called the holonomy. Under the
adiabatic evolution along a closed path in the parameter space, the state
vector will come back to itself up to an extra unitary phase factor, given
by the integration of path ordered exponential of the Berry holonomy along
that path.

We look for experimental quantities that can be realized from the structures
of (\ref{c_1}, \ref{c_2}). The field strengthes are
\begin{equation}
F_{ab}^{(1)ij}=-i\left\langle \partial _{a}\psi ^{i}\right. \left\vert
\partial _{b}\psi ^{j}\right\rangle +i\left\langle \partial _{b}\psi
^{i}\right. \left\vert \partial _{a}\psi ^{j}\right\rangle +i\left\langle
\psi ^{i}\right. \left\vert \partial _{a}\psi ^{l}\right\rangle \left\langle
\psi ^{l}\right. \left\vert \partial _{b}\psi ^{j}\right\rangle
-i\left\langle \psi ^{i}\right. \left\vert \partial _{b}\psi
^{l}\right\rangle \left\langle \psi ^{l}\right. \left\vert \partial _{a}\psi
^{j}\right\rangle ,
\end{equation}%
\begin{equation}
F_{ab}^{(2){\hat{\imath}}{\hat{\jmath}}}=-i\left\langle \partial _{a}\psi ^{{%
\hat{\imath}}}\right. \left\vert \partial _{b}\psi ^{{\hat{\jmath}}%
}\right\rangle +i\left\langle \partial _{b}\psi ^{{\hat{\imath}}}\right.
\left\vert \partial _{a}\psi ^{{\hat{\jmath}}}\right\rangle +i\left\langle
\psi ^{{\hat{\imath}}}\right. \left\vert \partial _{a}{\normalsize \psi }^{%
{\small \hat{l}}}\right\rangle ~\left\langle {\normalsize \psi }^{{\small
\hat{l}}}\right. \left\vert \partial _{b}\psi ^{{\hat{\jmath}}}\right\rangle
-i\left\langle \psi ^{{\hat{\imath}}}\right. \left\vert \partial _{b}\psi ^{%
{\small \hat{l}}}\right\rangle \left\langle \psi ^{{\small \hat{l}}}\right.
\left\vert \partial _{a}\psi ^{{\hat{\jmath}}}\right\rangle ,
\end{equation}%
$a,b~$label the parameter space, and $i,j~$label the state vector bases or
bands. The bracket denotes the inner product of wavefunctions in the Hilbert
space. One interesting situation is when $\xi _{a}=k_{a},$ $a=1,2,3$, and $%
\xi _{4}$ is an additional parameter, and this correspond to the momentum
space of 3D materials. Summation of the scripts are assumed in the notations.

Moreover, there are second Chern numbers $c_{2}^{(1)},c_{2}^{(2)}$ of these
two gauge fields,%
\begin{eqnarray}
c_{2}^{(1)} &=&\frac{1}{8\pi ^{2}}\int ~(\mathrm{tr}F^{(1)}\wedge F^{(1)})=%
\frac{1}{32\pi ^{2}}\int d^{4}\xi ~\epsilon _{abcd}(\mathrm{tr}%
F_{ab}^{(1)}F_{cd}^{(1)}),~  \label{fibration_01} \\
c_{2}^{(2)} &=&\frac{1}{8\pi ^{2}}\int ~(\mathrm{tr}F^{(2)}\wedge F^{(2)})=%
\frac{1}{32\pi ^{2}}\int d^{4}\xi ~\epsilon _{abcd}(\mathrm{tr}%
F_{ab}^{(2)}F_{cd}^{(2)}),  \label{fibration_02}
\end{eqnarray}%
where the integration is on the parameter space ($\xi _{1},\xi _{2},\xi
_{3},\xi _{4}$) and $d^{4}\xi =d\xi _{1}d\xi _{2}d\xi _{3}d\xi _{4}.$ The
second Chern numbers are
\begin{equation}
c_{2}^{(1)}=n+m,~~c_{2}^{(2)}=-n.
\end{equation}%
The non-trivial Chern numbers correspond to the non-trivial topology of the
band structure. It is an $S^{3}~$fibration of four-sphere. The $su(2)$
fibration on the 4d parameter space realizes a higher dimensional manifold.
The (\ref{fibration_01}, \ref{fibration_02}) can also be interpreted as
four-form magnetic monopole fluxes on the parameter space.

Realization and interpretation of the two $su(2)$'s and the parameter space
can be diverse, in condensed matter and atomic systems. This type of
fibration are still abstract. It can be mapped to band structures in
different possible systems. It has relevance to both electron band
structures and photon band structures. It is possible to have this band
structure in certain synthesized materials. The Chern numbers may correspond
to the number of edge states. One of the most interesting situation is $m=1$%
. To exhibit $c_{2}^{(1)}=n+1$,$~c_{2}^{(2)}=-n$, the material may have $%
n+1~ $right chiral edge states and $n~$left chiral edge states. In the
context of photonic crystals, it may have $n+1$ uni-directional right-moving
edge states, and $n$ uni-directional left-moving edge states. These
topological quantum numbers may manifest themselves in terms of the number
of edge states, and may contribute to conductivities and transport
properties of the materials.

In the situation that the fourth parameter $\xi _{4}$ can be integrated out,
the expressions can be reduced to Chern-Simons integrals, because of the
relation,%
\begin{equation}
\epsilon _{dabc}{\normalsize \nabla }_{d}[{\normalsize A}_{ij}^{a}%
{\normalsize \nabla }_{b}{\normalsize A}_{ji}^{c}+{\normalsize i}\frac{2}{3}%
{\normalsize A}_{il}^{a}{\normalsize A}_{lj}^{b}{\normalsize A}_{ji}^{c}]=%
\frac{1}{4}\epsilon _{abcd}(\mathrm{tr}F^{ab}F^{cd}).
\end{equation}%
Therefore the reduced integration keeps the information of the four
dimensional integral. In the case of momentum space $(k_{a},k_{b},k_{c})$
plus $\xi _{4}$,$~$the integration after reducing on $\xi _{4}~$is over the
Brillouin zone.

The four dimensional topological insulator can be characterized by Z
invariant. In 4D it may be characterized by $c_{2}\in \mathrm{Z.~}$So here
the system with $\mathrm{Z}\oplus \mathrm{Z}$ invariant may be mapped to
doubled topological insulators in four spatial dimensions. Upon a reduction
on the fourth parameter $\xi _{4},$ the system becomes a doubled three
dimensional topological insulators. The three dimensional topological
insulator have been characterized by $\mathrm{Z}_{2}$ invariant.\cite{Fu et
al, Moore Balents, Roy} The elements of $\mathrm{Z}_{2}$ correspond to odd
number or even number of Dirac points, which is related to the global
property of the Brillouin zone.\cite{Fu et al, Moore Balents, Roy} The odd
class are topological insulators, and the even class are normal insulators.

The fibration may be related to axion electrodynamics. An
effective axion term can be induced in several ways\cite{Qi et al,
Essin et al} in three spatial dimensions and one time dimension.
Similarly, in four spatial dimensions, there are effective
Chern-Simons action that can be induced\cite{Qi et al} in four
spatial dimensions and one time dimension. The Chern-Simons term
$A_{\kappa }\epsilon ^{\kappa \mu \nu \lambda \rho }\partial _{\mu
}A_{\nu }\partial _{\lambda }A_{\rho }$ is the one-loop effective
term arising from integrating the fermion loop and the coefficient
is given by the second Chern number in the momentum space of the
fermion. Up on reduction on $\kappa =x_{4}$ direction, $A_{\kappa
}$ becomes $\hat{\theta}$ field.

In three spatial dimensions and one time dimension, the systems
can effectively have axion term, or $\theta$ term, and have
magneto-electric effects,
\begin{equation}
S=\frac{e^{2}}{16\pi h}(c_{2}^{(1)}+c_{2}^{(2)})\int dtd^{3}x(\hat{\theta}%
\epsilon ^{\mu \nu \lambda \rho }\partial _{\mu }A_{\nu }\partial _{\lambda
}A_{\rho }),  \label{4D_S}
\end{equation}%
\begin{equation}
\theta =(c_{2}^{(1)}+c_{2}^{(2)})\hat{\theta}=m\hat{\theta},
\end{equation}%
where $A_{\mu }(x)~$is real space gauge field. When the system has
time-reversal symmetry, the time-reversal symmetry and gauge symmetry
require that $\theta =m\pi ,$ where $m$ is an integer. This term is
proportional to $\mathbf{E\cdot B,}$ and thus this will give
magneto-electric polarization of the material.

If we consider the interface between two materials with different $\hat{%
\theta},$ in which case there is$~\hat{\theta}_{1}~$for one material
extending along $z<0$, and $\hat{\theta}_{2}~$for the other material
extending along $z>0.~$There is a jump $\hat{\theta}_{1}-\hat{\theta}%
_{2}=\Delta \hat{\theta}$ across the two sides of the interface. The
time-reversal symmetry can be broken on this interface.\cite{Qi et al, Essin
et al} We can use the integration by parts $(\hat{\theta})\partial
_{z}A_{x}\partial _{t}A_{y}=-A_{x}[(\partial _{z}\hat{\theta})\partial
_{t}A_{y}]~$up to total derivatives. Since $A_{x}~$is~coupled to the current
$j_{x}~$via $A_{x}j_{x},$ then at the interface we have that the induced
current is%
\begin{eqnarray}
j_{x} &=&\sigma _{xy}E_{y}, \\
j_{x} &=&j_{x}^{(1)}+j_{x}^{(2)},~\ ~\ ~j_{x}^{(1)}=\sigma
_{xy}^{(1)}E_{y},~~~~~j_{x}^{(2)}=\sigma _{xy}^{(2)}E_{y}. \\
j_{x}^{(1)} &=&\frac{e^{2}}{h}\frac{(\hat{\theta}_{1}-\hat{\theta}_{2})}{%
2\pi }E_{y}c_{2}^{(1)}=(n+m)\frac{e^{2}}{h}\frac{(\hat{\theta}_{1}-\hat{%
\theta}_{2})}{2\pi }E_{y}, \\
j_{x}^{(2)} &=&\frac{e^{2}}{h}\frac{(\hat{\theta}_{1}-\hat{\theta}_{2})}{%
2\pi }E_{y}c_{2}^{(2)}=-n\frac{e^{2}}{h}\frac{(\hat{\theta}_{1}-\hat{\theta}%
_{2})}{2\pi }E_{y}.
\end{eqnarray}%
The two conductivities have opposite signs. From the expression of the
second Chern numbers, the conductance can be expressed as
\begin{eqnarray}
\sigma _{xy}^{(1)} &=&\frac{1}{4\pi ^{2}}\int d^{3}k[\left\langle \psi
^{i}\right. \left\vert \partial _{\mu }\psi ^{j}\right\rangle \left\langle
\partial _{\nu }\psi ^{j}\right. \left\vert \partial _{\lambda }\psi
^{i}\right\rangle +\frac{2}{3}\left\langle \psi ^{i}\right. \left\vert
\partial _{\mu }\psi ^{j}\right\rangle \left\langle \psi ^{j}\right.
\left\vert \partial _{\nu }\psi ^{l}\right\rangle \left\langle \psi
^{l}\right. \left\vert \partial _{\lambda }\psi ^{i}\right\rangle ]\frac{%
e^{2}}{h}\frac{(\hat{\theta}_{1}-\hat{\theta}_{2})}{2\pi }\epsilon ^{\mu \nu
\lambda },  \notag \\
\sigma _{xy}^{(2)} &=&\frac{1}{4\pi ^{2}}\int d^{3}k[\left\langle \psi
^{\alpha }\right. \left\vert \partial _{\mu }\psi ^{\beta }\right\rangle
\left\langle \partial _{\nu }\psi ^{\beta }\right. \left\vert \partial
_{\lambda }\psi ^{\alpha }\right\rangle +\frac{2}{3}\left\langle \psi
^{\alpha }\right. \left\vert \partial _{\mu }\psi ^{\beta }\right\rangle
\left\langle \psi ^{\beta }\right. \left\vert \partial _{\nu }\psi ^{\gamma
}\right\rangle \left\langle \psi ^{\gamma }\right. \left\vert \partial
_{\lambda }\psi ^{\alpha }\right\rangle ]\frac{e^{2}}{h}\frac{(\hat{\theta}%
_{1}-\hat{\theta}_{2})}{2\pi }\epsilon ^{\mu \nu \lambda },  \notag \\
&&
\end{eqnarray}%
where $\int d^{3}k=\int dk_{x}dk_{y}dk_{z},~$and where there is summation in
the $i,j,\alpha ,\beta ~$labels of different occupied bands. The integration
of the second Chern class over $I\times BZ$ reduces to the integration of
two Chern-Simons forms on $BZ$.

The fibration is also related to band touching phenomena. There are
nontrivial topological structures of occupied bands. There are also other
unoccupied bands. We can diagonalize the Hamiltonian in the system, and it
can take the form%
\begin{equation}
H(\mathbf{k})=E_{1}(\mathbf{k})\sum_{i=1,2}\left\vert i,\mathbf{k}%
\right\rangle \left\langle i,\mathbf{k}\right\vert +E_{2}(\mathbf{k}%
)\sum_{\alpha =1,2}\left\vert \alpha ,\mathbf{k}\right\rangle \left\langle
\alpha ,\mathbf{k}\right\vert +\sum_{\gamma }E_{\gamma }(\mathbf{k}%
)\left\vert \gamma ,\mathbf{k}\right\rangle \left\langle \gamma ,\mathbf{k}%
\right\vert .
\end{equation}%
We have that $E_{1},E_{2}~$are energy eigenvalues of the two double bands.
Those are occupied bands. We have written it in orthonormal basis. $%
E_{\gamma }$ correspond to other unoccupied bands. The second Chern numbers
of the occupied bands are $n+m,-n$ respectively. We propose that there may
exist band structures with associated geometric phase that realize the
exotic spheres.

When we tune the parameters in the model describing the material, the band
structures can be deformed and changed. As long as there is no band touching
that occurs during the tuning, the individual Chern number of each band
remain unchanged, since it is topologically invariant. If when tuning the
parameters, the band touching happens, then the Chern numbers of two bands
that touch can change their individual Chern numbers. In these situations,
when tuning the parameters, the two bands first touch and then split. If we
view the geometric phase gauge field as a fiber bundle over the parameter
space, for example the three dimensional momentum space plus an additional
parameter, then the transferring of Chern numbers during band-touching is a
topology change of that fiber bundle. The base space is the parameter space,
and the fiber is the geometric phase gauge field. The Chern number of the
band may correspond to the number of edge states.

Consider the effective Hamiltonian of the two double bands,
\begin{equation}
H(\mathbf{k;}{\normalsize \xi })=\left( \frac{E_{1}(\mathbf{k;}{\normalsize %
\xi })+E_{2}(\mathbf{k{\normalsize ;}}{\normalsize \xi })}{2}\right) \text{I}%
_{4\times 4}-\left( \frac{E_{1}(\mathbf{k;}{\normalsize \xi })-E_{2}(\mathbf{%
k{\normalsize ;}}{\normalsize \xi })}{2}\right) \vec{\Gamma}\cdot \vec{\omega%
}(\mathbf{k;}{\normalsize \xi }),
\end{equation}%
where we write it in terms of $4\times 4~$Hamiltonian $H(\mathbf{k;}%
{\normalsize \xi })$, and where $\vec{\omega}\cdot \vec{\omega}=1,$ and $%
\Gamma ~$denotes Gamma matrices$~$and ${\normalsize \xi }$ here denotes a
parameter in the model Hamiltonian, for example spin-orbit coupling. If the
band touching happens at a point near ($\mathbf{\hat{k}};\hat{\xi}$), we
draw a surface $\Sigma _{3}~$enclosing that point. The Chern number transfer
between the two bands is therefore
\begin{equation}
{\hat{n}}=\frac{1}{8\pi ^{2}}\int_{\Sigma _{3}}\left\langle \vec{\omega}%
\right. \left\vert \vec{\omega}d\vec{\omega}\wedge d\vec{\omega}\wedge d\vec{%
\omega}\right\rangle .
\end{equation}%
This is the Chern number that is transferred between the two bands.

Starting from the state corresponding to $c_{2}^{(1)}=n+m,~c_{2}^{(2)}=-n,$
we can tune the parameters in the model, for example, the spin-orbit
coupling, or strain, or magnetic field, so as to change the band structure
to make the band touching point between the two bands happen. The two bands
then split by tuning these parameters. When the bands touch and then split,
they transfer Chern number $\Delta c_{2}={\hat{n}},$ and then the state
becomes $c_{2}^{(1)}=n^{\prime }+m,~c_{2}^{(2)}=-n^{\prime },$ in which $%
n^{\prime }=n+{\hat{n}.~}$From the point of view of instantons on the base
space of the exotic sphere fibration, this transition between the two bands
correspond to the transfer of ${\hat{n}}$ instantons between the two $su(2)$
gauge fields.~This also corresponds to the topological change transition
between exotic sphere fibrations with different $n$.~The same procedure of
band touching and Chern number transfer can be performed many times. This
generates different fibrations with $c_{2}^{(1)}=n+m,~c_{2}^{(2)}=-n$, and
exotic spheres with $c_{2}^{(1)}=n+1,~c_{2}^{(2)}=-n$. The change of Chern
numbers can be mapped to the topology change of the fiber bundle. These
transitions may relate different $n$ with fixed $m$.

Those states with fixed $m=1$, fixed $n$ belong to the same diffeomorphism
class, and they may correspond to the situations that the topology of the
band structure does not change, and in particular without band touching and
Chern number transfer. Those with fixed $m=1$, but different $n$, are in the
same homeomorphism class, but for different $n$ (mod 28) are not in the same
diffeomorphism class, and they may be non-adiabatically connected by band
touchings.

Those with different $m$ are not in the same homeomorphism class and may not
be non-adiabatically connected. Since the two bands touch and split, the
total Chern numbers of the two occupied bands, which is $m$, is
topologically invariant and a conserved quantity of the total system. There
are changes of the individual Chern numbers of each band, and there is a
Chern number transfer between the two bands.

Many electron systems can be described by effective Hamiltonian that is
quadratic in the Fermi field. In some situations there may contain terms
that are quartic in the Fermi field. The quartic terms can give radiative
loop corrections to the quadratic terms. In some situations that a mean
field theory can be applied, and quartic term may be substituted by a
quadratic term by a mean field approximation.

The system may be related to diverse systems, for example, topological
insulators, quantum hall systems, models of Weyl semimetals, semiconductors,
photonic crystals, and cold atom systems.

\section{Theoretical doubled Topological insulator model}


\label{sec_ topological insulators}

The 3D topological insulator materials include for example the \textrm{Bi}$%
_{2}\mathrm{Te}_{3},~$\textrm{Bi}$_{2}\mathrm{Se}_{3},$ \textrm{Sb}$_{2}%
\mathrm{Te}_{3},~$and \textrm{Bi}$_{1-x}\mathrm{Sb}_{x}~$(for example $%
\mathrm{Bi}_{0.9}\mathrm{Sb}_{0.1}$).$~$The alloy $\mathrm{Bi}_{0.9}\mathrm{%
Sb}_{0.1}$ has five Fermi crossing points, and its theoretical model is more
complicated than the situations of \textrm{Bi}$_{2}\mathrm{Te}_{3},~$\textrm{%
Bi}$_{2}\mathrm{Se}_{3},$ \textrm{Sb}$_{2}\mathrm{Te}_{3},$ which have a
single Dirac point. The topological insulator materials can be known by
measuring whether there is odd or even number of Dirac points, which is
related to the global property of the Brillouin zone.

The 3D Topological Insulator materials, for example\cite{Xia etal, Zhang
etal, Chen Y L etal} \textrm{Bi}$_{2}\mathrm{Te}_{3},~$\textrm{Bi}$_{2}%
\mathrm{Se}_{3},$ \textrm{Sb}$_{2}\mathrm{Te}_{3},$ can be described by a
simplified effective $4\times 4~$model Hamiltonian, near the level-crossing
point. They have large bulk gap of order ($1\sim 3$)$\times 10^{-1}~$eV.~The
model with $\mathrm{Se}1$-\textrm{Bi}$1$-$\mathrm{Se}2$-\textrm{Bi}$%
1^{\prime }$-$\mathrm{Se}1^{\prime }$ crystal structure have been studied in
detail.\cite{Zhang etal} For example, for a \textrm{Bi}$_{2}\mathrm{Se}_{3}$
crystal, in the effective model of the four band model, the basis of the
state vector is
\begin{equation}
|P1_{z}^{+},\mathbf{\uparrow }\rangle ,~|P2_{z}^{-},\mathbf{\uparrow }%
\rangle ,~|P1_{z}^{+},\mathbf{\downarrow }\rangle ,~|P2_{z}^{-},\mathbf{%
\downarrow }\rangle
\end{equation}%
where $P1_{z}^{+},P2_{z}^{-}~$are two $p~$orbitals in the situations in a
\textrm{Bi}$_{2}\mathrm{Se}_{3}$ crystal. Near the level-crossing point, the
two bands touch each other. In this effective model, the four basis are from
two orbitals and two spins.$~$Let's abstract the state vector into%
\begin{equation}
\left\vert +,\mathbf{\uparrow }\right\rangle ,~\left\vert -,\mathbf{\uparrow
}\right\rangle ,~\left\vert +,\mathbf{\downarrow }\right\rangle ,~\left\vert
-,\downarrow \right\rangle .~
\end{equation}

We may introduce another $su(2)$ pseudospin space, in which we enlarge the $%
4\times 4~$model Hamiltonian and 4 vector into $8\times 8~$model Hamiltonian
and 8 vector
\begin{equation}
\left\vert +,1,\mathbf{\uparrow }\right\rangle ,~\left\vert -,1,\mathbf{%
\uparrow }\right\rangle ,~\left\vert +,1,\mathbf{\downarrow }\right\rangle
,~\left\vert -,1,\downarrow \right\rangle ,~\left\vert +,2,\mathbf{\uparrow }%
\right\rangle ,~\left\vert -,2,\mathbf{\uparrow }\right\rangle ,~\left\vert
+,2,\mathbf{\downarrow }\right\rangle ,~\left\vert -,2,\downarrow
\right\rangle .  \label{state_}
\end{equation}%
The pseudospin refers to the labels 1 and 2 in (\ref{state_}). There are
potentially many different kinds of 3D topological insulator materials. It
is in principle possible in the future to consider Interpenetrating Lattices
of two topological insulator materials, or Interpenetrating Lattices of one
topological insulator and one normal insulator material experimentally.
Here, we only discuss it theoretically. The model Hamiltonian of the $%
8\times 8~$model can be described as
\begin{equation}
H(\mathbf{k})=\epsilon _{0}(\mathbf{k,}{\normalsize \xi })\text{I}_{8\times
8}+\left[
\begin{array}{cc}
g(\mathbf{k,}{\normalsize \xi }) & 0 \\
0 & -g(\mathbf{k,}{\normalsize \xi })%
\end{array}%
\right] \otimes \text{I}_{4\times 4}+\left[
\begin{array}{cc}
d^{(1)}(\mathbf{k,}{\normalsize \xi })n_{a}^{(1)}(\mathbf{k,}{\normalsize %
\xi }) & 0 \\
0 & d^{(2)}(\mathbf{k,}{\normalsize \xi })n_{a}^{(2)}(\mathbf{k,}%
{\normalsize \xi })%
\end{array}%
\right] \otimes \Gamma _{a}.
\end{equation}%
$n_{a}^{(1)}(\mathbf{k,}{\normalsize \xi })$ and $n_{a}^{(2)}(\mathbf{k,}%
{\normalsize \xi })$ are unit-norm vectors mapped from $(\mathbf{k,}%
{\normalsize \xi })$ space. It is a $8\times 8~$model, doubled from $4\times
4~$model.\cite{Qi et al} The model is analogous to doubled topological
insulators, and have two second Chern numbers%
\begin{eqnarray}
c_{2}^{(1)} &=&\frac{3}{8\pi ^{2}}\int d^{3}kd{\normalsize \xi }\epsilon
^{abcde}n_{a}^{(1)}\partial _{k_{1}}n_{b}^{(1)}\partial
_{k_{2}}n_{c}^{(1)}\partial _{k_{3}}n_{d}^{(1)}\partial _{{\normalsize \xi }%
}n_{e}^{(1)}=m+n,~~ \\
c_{2}^{(2)} &=&\frac{3}{8\pi ^{2}}\int d^{3}kd{\normalsize \xi }\epsilon
^{abcde}n_{a}^{(2)}\partial _{k_{1}}n_{b}^{(2)}\partial
_{k_{2}}n_{c}^{(2)}\partial _{k_{3}}n_{d}^{(2)}\partial _{{\normalsize \xi }%
}n_{e}^{(2)}=-n.
\end{eqnarray}%
These integral representations also give the second Chern numbers.

In that case, the boundary states may compose of $m+n$ right chiral fermion
modes, and $n$ left chiral fermion modes. The Hamiltonian density of these
states in momentum space may be expressed as
\begin{equation}
H(\mathbf{k})=\sum_{i=1,...,m+n}\hbar v_{i}\psi _{i}^{\dagger }(\mathbf{%
\sigma }\cdot \mathbf{k})\psi _{i}+\sum_{j=1,...,n}\hbar v_{j}\psi
_{j}^{\dagger }(-\mathbf{\sigma }\cdot \mathbf{k})\psi _{j}.
\end{equation}

In general this fibration (\ref{fibration_01}, \ref{fibration_02}) may be
mapped to $m+n$ right chiral modes, and $n$ left chiral modes. The total
helicity number is $m$.

It can be realized in Interpenetrating Lattices of two kinds of insulator
materials. In the context of topological insulators in three spatial
dimensions, the situation with the odd number of edge states is
topologically robust. Since one can perform perturbations to the system and
a pair of Dirac cones can be coupled and then gapped after
re-diagonalization of the Hamiltonian. For the odd number of Dirac cones,
such perturbations will always leave at least one Dirac cone un-gapped.

For $m=2$, it can be realized in Interpenetrating Lattices of two kinds of
topological insulator materials. For $m=1$, it can be realized in
Interpenetrating Lattices of one kind of topological insulator material, and
one kind of normal insulator material.

Because of the $u(1)$ symmetry, the electric current in the system is
exactly conserved current. Similar to the discussion in section \ref{sec_
geometric phase band chern number}, the electric charge current is
\begin{eqnarray}
j_{x}^{(1)} &=&\frac{e^{2}}{h}\frac{(\hat{\theta}_{1}^{(1)}-\hat{\theta}_{2})%
}{2\pi }c_{2}^{(1)}E_{y}, \\
j_{x}^{(2)} &=&\frac{e^{2}}{h}\frac{(\hat{\theta}_{1}^{(2)}-\hat{\theta}_{2})%
}{2\pi }c_{2}^{(2)}E_{y},
\end{eqnarray}%
at the interface between the Interpenetrating Lattices of the two materials
with $\hat{\theta}_{1}^{(1)}$, $\hat{\theta}_{1}^{(2)}$ respectively, and
another material with $\hat{\theta}_{2}.$

In the context of Interpenetrating Lattices, the enlargement from $4\times
4~ $to $8\times 8$ is due to two types of interweaving lattice site L$^{%
{\small (1)}}$, L$^{{\small (2)}}$. Measurement associated with particular
sublattice L$^{{\small (1)}}$ or L$^{{\small (2)}}$ selects the
corresponding Chern number.

\section{Photonic crystals}


\label{sec_ photonic crystals}

We may connect these fibrations to electron band and photon band. The
photonic bands are parallel and similar to electronic bands. Photonic
crystal with bulk band-gap, and gapless edge modes are in some aspects
similar to topological insulator. It can have several bulk band-gaps. It may
have special edge states. Since the material has bulk band-gap for the
photon, it will forbid the bulk transmission of the photons in certain range
of frequencies, for example $\omega _{2}<\omega <\omega _{1}$. There can be
surface band that are within the band-gap region of the bulk bands.

There are many ways to engineer photon bands in photonic crystals, and there
are typically many closely-spaced bands. So there are many possibilities to
have several Chern numbers. There are Dirac points near band touching
points. The photon bands also have geometric phases. The 2D photonic
crystals can be made by periodic array of cylinders of dielectric medium,
with lattice structure in $x,y$ directions. The geometric phase of 2D
photonic crystal can be defined and its first Chern number is given by
integration of the field strength of the Berry phase gauge field in the 2D
momentum space ($k_{x},k_{y}$), for example.\cite{H R}

The 3D photonic crystals (PhC) and 2D photonic crystals (PhC) have a
difference that the 2D photonic crystals have extra translational symmetry
in $z$ direction. The 3D photonic crystals can be made by 3D periodic arrays
(or lattices) of dielectric spheres, or alternatively by 3D periodic arrays
(lattices) of air holes in dielectric medium, or by 3D meshes of dielectric
medium. There can be a limit that the lattice spacing along $z$ direction is
much smaller than the lattice spacings in $x,y$ directions, and under such
limit it cross over to 2D system. One can also define a conductivity of edge
modes of photons in 3D.

The band topology can be realized also in three dimensional photonic
crystals. The geometric phase can also be similarly expressed
\begin{equation}
\mathcal{A}_{ij}^{a}(\mathbf{k})=\mathrm{Im}\left( \frac{\left( \boldsymbol{u%
}_{i}(\mathbf{k}),\tilde{{\normalsize B}}^{-1}(\omega _{j}(\mathbf{k}%
))\nabla _{k_{a}}\boldsymbol{u}_{j}(\mathbf{k})\right) }{\sqrt{\left(
\boldsymbol{u}_{i}(\mathbf{k}),{\normalsize \tilde{B}}^{-1}(\omega _{i}(%
\mathbf{k}))\boldsymbol{u}_{i}(\mathbf{k})\right) \left( \boldsymbol{u}_{j}(%
\mathbf{k}),{\normalsize \tilde{B}}^{-1}(\omega _{j}(\mathbf{k}))\boldsymbol{%
u}_{j}(\mathbf{k})\right) }}\right) .
\end{equation}%
\begin{equation}
\mathcal{A}_{\alpha \beta }^{a}(\mathbf{k})=\mathrm{Im}\left( \frac{\left(
\boldsymbol{u}_{\alpha }(\mathbf{k}),\tilde{{\normalsize B}}^{-1}(\omega
_{\beta }(\boldsymbol{k}))\nabla _{k_{a}}\boldsymbol{u}_{\beta }(\mathbf{k}%
)\right) }{\sqrt{\left( \boldsymbol{u}_{\alpha }(\mathbf{k}),{\normalsize
\tilde{B}}^{-1}(\omega _{\alpha }(\mathbf{k}))\boldsymbol{u}_{\alpha }(%
\mathbf{k})\right) \left( \boldsymbol{u}_{\beta }(\mathbf{k}),{\normalsize
\tilde{B}}^{-1}(\omega _{\beta }(\mathbf{k}))\boldsymbol{u}_{\beta }(\mathbf{%
k})\right) }}\right) .
\end{equation}%
The round bracket denotes contraction of spatial components of the
field variables.\cite{H R} Here we include non-abelian geometric
phases. The $i,j$ labels a double band and takes values 1 or 2.
The $\alpha ,\beta ~$labels a different double band and takes
values 1 or 2. We have assumed almost degeneracy $\omega
_{i}\simeq \omega _{j}$,$~$and $\omega _{\alpha }\simeq \omega
_{\beta }.$ We assume that there are two almost doubly degenerate
bands in the band structure.

$B^{-1}(r,\omega )$ is an $6\times 6$ block-diagonal
permittivity-permeability tensor
\begin{equation}
B^{-1}(r,\omega )=\left[
\begin{array}{cc}
\epsilon _{ab}(r,\omega ) & 0 \\
0 & \mu _{ab}(r,\omega )%
\end{array}%
\right] .
\end{equation}%
${\normalsize \tilde{B}}^{-1}(\omega )~$has$~$taken into account frequency
dependence,\cite{H R}~%
\begin{equation}
{\normalsize \tilde{B}}^{-1}(r,\omega )=B^{-1}(r,\omega )+\omega \partial
_{\omega }B^{-1}(r,\omega ).
\end{equation}

The $\epsilon _{ab}(r,\omega ),~\mu _{ab}(r,\omega )~$are $3\times 3~$%
permittivity tensor and permeability tensor, and they generally have
off-diagonal components. ${\normalsize \tilde{B}}^{-1}(\omega (\mathbf{k}))~$%
is a nontrivial tensor due to the dielectric medium, and they have frequency
dependence.\cite{H R}

$~$The$~\boldsymbol{u}_{i}(\mathbf{k},r)e^{i\mathbf{k}\cdot r}$ is the Bloch
state of the 6-component complex vector $(\tilde{E}_{i}(\mathbf{k},r),\tilde{%
H}_{i}(\mathbf{k},r)),$ of the electromagnetic fields of the normal mode
with momentum vector $\mathbf{k}$ and frequency $\omega _{i}(\mathbf{k})$.

One can define a Chern-Simons integral,
\begin{equation}
I=\frac{1}{4\pi }\int dk_{x}dk_{y}dk_{z}\epsilon _{abc}[\mathcal{A}%
_{ij}^{a}\nabla _{k_{b}}\mathcal{A}_{ji}^{c}+i\frac{2}{3}\mathcal{A}_{il}^{a}%
\mathcal{A}_{lj}^{b}\mathcal{A}_{ji}^{c}].  \label{integral_}
\end{equation}%
The integral is in the 3D momentum space ($k_{x},k_{y},k_{z}$) of the 3D
photonic crystal, and summation of the scripts are assumed in the notations.

Because of the relation,
\begin{equation}
\epsilon _{dabc}\nabla _{d}[\mathcal{A}_{ij}^{a}\nabla _{b}\mathcal{A}%
_{ji}^{c}+i\frac{2}{3}\mathcal{A}_{il}^{a}\mathcal{A}_{lj}^{b}\mathcal{A}%
_{ji}^{c}]=\frac{1}{4}\epsilon _{abcd}(\mathrm{tr}F^{ab}F^{cd})
\end{equation}%
the four dimensional integral of
\begin{equation}
\frac{1}{4}\int d\xi _{a}d\xi _{b}d\xi _{c}d\xi _{d}~\epsilon _{abcd}(%
\mathrm{tr}F^{ab}F^{cd})
\end{equation}%
can be reduced to three dimensional integral of
\begin{equation}
\int d\xi _{a}d\xi _{b}d\xi _{c}[\mathcal{A}_{ij}^{a}\nabla _{b}\mathcal{A}%
_{ji}^{c}+i\frac{2}{3}\mathcal{A}_{il}^{a}\mathcal{A}_{lj}^{b}\mathcal{A}%
_{ji}^{c}].
\end{equation}%
Therefore the Chern-Simons integral has the information of the
four dimensional integral that is associated with the $c_{2}$.

For particularly engineered 3D photonic crystals, there could be two second
Chern numbers $c_{2}^{(1)},c_{2}^{(2)}$, whose values may correspond to the
numbers of right-moving and left-moving boundary states. This type of band
structure can also be realized in Interpenetrating Lattices of two photonic
crystal materials PhC and PhC$^{\prime }$, experimentally. In the latter
case, $c_{2}^{(1)}$,$~c_{2}^{(2)}$ correspond to the two materials
respectively. The Chern numbers may correspond to the number of boundary
states, or uni-directional one-way propagating states on the boundary. This
is independent of boson or fermion statistics. Photons can have right
circular polarization and left circular polarization. In this context, the
photon's left or right polarization pattern of the boundary photon states
would be interesting physical observable.

There can be a limit that the lattice spacing along $z$ direction is much
smaller than the lattice spacings in $x,y$ directions, and the 3D system can
crossover to the 2D system. The crossover relation between 3D TI and 2D TI
in some aspects may be similar to the relation between 3D PhC and 2D PhC.

\section{Heterostructure of TI/NI/TI/NI$^{\prime}$}


\label{sec_ heterostructure TI/NI/TI/NI'}

In this section we discuss heterostructures of periodic units of TI/NI/TI/NI$%
^{\prime }$ materials. The periodic heterostructure of TI/NI has
been devised.\cite{BuBa, BuHoBa} The TI/NI stands for Topological
Insulator/Normal Insulator. In periodic TI/NI model,\cite{BuBa,
BuHoBa} there are two kinds of interfaces, NI/TI and TI/NI. In the
periodic TI/NI/TI/NI$^{\prime }$
model, there are four kinds of surfaces, NI/TI, TI/NI$^{\prime }$, NI$%
^{\prime }$/TI, TI/NI. These materials are arranged along the $z$
direction layer by layer, from top to bottom direction. Here we
make the normal insulators NI and NI$^{\prime }$ on the two sides
of the same topological insulator to be different. The difference
of NI and NI$^{\prime }$ introduces another ${\normalsize su(2)}$
space, the $\rho $-space. We make the parameters of two kinds of
TI/NI, TI/NI$^{\prime }$ junctions to be different, so in each
periodic unit, there are four materials. There is experimental
method to make the heights of two normal insulators to be the
same, while making the tunnelings of the surface electrons across
the two kinds of normal insulators to be different.\ This type of
structure can be experimentally performed by many layers of
periodic heterostructure of thin films.

In each periodic unit, there are TI, NI, TI, NI$^{\prime }~$structures.
There are surface electrons in\ upper surface and lower surface of the TI
materials (in $x,y$ directions). The model Hamiltonians are built from the
states on the interfaces, from the surface states of TI's. Adding the normal
insulator materials can perturb the surface Hamiltonian by adding tunneling
terms. The periodic structure is along the $z$ direction. The periodic
structure then make the states become bulk states of the engineered
structure.

The model Hamiltonian is%
\begin{eqnarray}
H &=&\sum_{\boldsymbol{k}_{\perp },i,j}\left[ (\hbar v_{F}\tau ^{z}(\hat{z}%
\times \mathbf{\sigma })\cdot \mathbf{k}_{\perp }+\Delta _{s}\tau
^{x}+b_{z}\sigma ^{z}+b_{x}\sigma ^{x})(\delta _{i,2j}+\delta
_{i,2j+1})\right.  \notag \\
&+&\left. \frac{1}{2}\tau ^{+}(\Delta _{1}\delta _{i,2j+1}+\Delta _{2}\delta
_{i,2j})+\frac{1}{2}\tau ^{-}(\Delta _{1}\delta _{i,2j}+\Delta _{2}\delta
_{i,2j+1})\right] c_{\boldsymbol{k}_{\perp }i}^{\dagger }(c_{\boldsymbol{k}%
_{\perp }2j+1}+c_{\boldsymbol{k}_{\perp }2j}).  \label{H_NI}
\end{eqnarray}%
This is generalized from the model$~$Hamiltonian of the TI/NI model with two
structures in each periodic unit.\cite{BuBa} $\tau ^{x},\tau ^{y},\tau ^{z}~$%
are Pauli matrices, which act on the pseudospin space of upper and lower
surfaces, and $\tau ^{+}=\tau ^{x}+i\tau ^{y},\tau ^{-}=\tau ^{x}-i\tau ^{y}$%
.$~$The $\mathbf{k}_{\perp }$ is the momentum in $x,y$ directions. The $i$
and $2j,2j+1$ label different topological insulator layers. The Hamiltonian (%
\ref{H_NI}) can describe the periodic structure of topological insulators
stacked together with normal insulators NI and NI$^{\prime }$ in between,
separating the topological insulators. The $b_{z}\sigma ^{z}+b_{x}\sigma
^{x} $ term in (\ref{H_NI}) describes spin splitting of the surface states,
that can be induced by doping each TI layer with magnetic impurities. $%
\Delta _{s} $ describes the tunneling between the two surfaces of the same
topological insulator. $\Delta _{1}~$and $\Delta _{2}~$describe the
tunneling between the surfaces of two nearby topological insulators through
the material in the middle, which are the NI and NI$^{\prime }~$%
respectively. The $\Delta _{1}~$and $\Delta _{2}$ parameters have different
sizes, and can be the same under the limit $\Delta _{2}/\Delta
_{1}\rightarrow 1$. The spacing of the periodic structure is $d$,~and the
total number of the periodic units is $N.$ The parameters for TI materials
are surface Fermi velocity $v_{F}$,~$b_{z},b_{x}~$and tunneling $\Delta
_{s}.~$The parameters for NI materials are the tunnelings $\Delta
_{1},\Delta _{2}$.$~$The $b_{z}\sigma ^{z}+b_{x}\sigma ^{x}$ is a
time-reversal symmetry breaking term.

There are several differences between the configurations here and the
configurations in previous discussion.\cite{BuBa, BuHoBa} In the
configurations there,\cite{BuBa, BuHoBa} there are two structures, the
topological insulator and normal insulator in each unit. Here, there are
four structures in each periodic unit, the TI, NI, TI, NI$^{\prime }$. Here,
we turn on the magnetic term in both $z$ and $x$ directions.

Making a Fourier transformation along the $z$ direction,
\begin{equation}
c_{\boldsymbol{k}_{\perp },l}^{\dagger }=\frac{1}{\sqrt{N}}\sum_{k_{z}}c_{%
\boldsymbol{k}}^{\dagger }e^{-ik_{z}ld},  \label{transform_}
\end{equation}%
where $N$ is the total number of the periodic units, the 3D momentum-space
Hamiltonian is a $8\times 8~$Hamiltonian,%
\begin{equation}
H=\sum_{\boldsymbol{k}}c_{\boldsymbol{k}}^{\dagger }H(\mathbf{k})c_{%
\boldsymbol{k}},
\end{equation}%
\begin{eqnarray}
H(\mathbf{k}) &=&  \notag \\
&&{\normalsize \left[
\begin{array}{cccc}
\varepsilon {\small +b}_{z}{\small \sigma }^{z}{\small +b}_{x}{\small \sigma
}^{x} & {\small \Delta }_{s}\text{I}_{\sigma } & {\small 0} & {\small \Delta
}_{2}{\small e}^{ik_{z}d}\text{I}_{\sigma } \\
{\small \Delta }_{s}\text{I}_{\sigma } & {\small -\varepsilon +b}_{z}{\small %
\sigma }^{z}{\small +b}_{x}{\small \sigma }^{x} & {\small \Delta }_{1}\text{I%
}_{\sigma } & {\small 0} \\
{\small 0} & {\small \Delta }_{1}\text{I}_{\sigma } & \varepsilon {\small +b}%
_{z}{\small \sigma }^{z}{\small +b}_{x}{\small \sigma }^{x} & {\small \Delta
}_{s}\text{I}_{\sigma } \\
{\small \Delta }_{2}{\small e}^{-ik_{z}d}\text{I}_{\sigma } & {\small 0} &
{\small \Delta }_{s}\text{I}_{\sigma } & {\small -\varepsilon +b}_{z}{\small %
\sigma }^{z}{\small +b}_{x}{\small \sigma }^{x}%
\end{array}%
\right] ,}
\end{eqnarray}%
where $\varepsilon ={\small \hbar v}_{F}{\small (\hat{z}\times }\mathbf{%
\sigma }{\small )\cdot }\mathbf{k}.$\vspace{1pt} The Hamiltonian here is $%
8\times 8.~$Because another ${\normalsize su(2)}$ $\rho $-space is
introduced when $\Delta _{2}\neq \Delta _{1}$, the Hamiltonian is enlarged
from $4\times 4~$to $8\times 8.$

We introduce a $su(2)$ space, $\rho $-space,~where $\rho ^{x},\rho
^{y},\rho ^{z}~$are Pauli matrices. The Hamiltonian can be
expressed as
\begin{eqnarray}
H(\mathbf{k}) &=&\left[ \hbar v_{F}\tau ^{z}\otimes ({\hat{z}}\times \mathbf{%
\sigma })\cdot \mathbf{k}+\Delta _{s}\tau ^{x}\otimes \text{I}_{\sigma }+%
\text{I}_{\tau }\otimes (b_{z}\sigma ^{z}+b_{x}\sigma ^{x})\right] \otimes
\text{I}_{\rho }  \notag \\
&&+\frac{1}{2}[\tau ^{+}\Delta _{1}+\tau ^{-}\Delta _{2}{\small e}%
^{ik_{z}d}]\rho ^{+}\frac{1}{2}\otimes \text{I}_{\sigma }+\frac{1}{2}[\tau
^{-}\Delta _{1}+\tau ^{+}\Delta _{2}{\small e}^{-ik_{z}d}]\rho ^{-}\frac{1}{2%
}\otimes \text{I}_{\sigma },
\end{eqnarray}%
where $\rho ^{+}=\rho ^{x}+i\rho ^{y},\rho ^{-}=\rho ^{x}-i\rho ^{y},~$and I$%
_{\rho }$ is the identity in $\rho $-space.

In the special case, when $b_{x}=0,~$making the transformation $\tau ^{\pm
}\rightarrow \tau ^{\pm }\sigma ^{z},\sigma ^{\pm }\rightarrow \sigma ^{\pm
}\tau ^{z},$ we find%
\begin{eqnarray}
H(\mathbf{k}) &=&\left[ \hbar v_{F}({\hat{z}}\times \mathbf{\sigma })\cdot
\mathbf{k}+b_{z}\sigma ^{z}\right] ~\text{I}_{\tau }\otimes \text{I}_{\rho
}+\sigma ^{z}\Sigma , \\
\Sigma &=&[\Delta _{s}\tau ^{x}\otimes \text{I}_{\rho }+\frac{1}{2}\tau
^{+}(\Delta _{1}\rho ^{+}+\Delta _{2}{\small e}^{-ik_{z}d}\rho ^{-})\frac{1}{%
2}+\frac{1}{2}\tau ^{-}(\Delta _{1}\rho ^{-}+\Delta _{2}{\small e}%
^{ik_{z}d}\rho ^{+})\frac{1}{2}].
\end{eqnarray}%
$\Sigma $ does not contain $\sigma $,$~$but only $\tau ,\rho ~$operators, so
it commutes with $\left[ \hbar v_{F}({\hat{z}}\times \mathbf{\sigma })\cdot
\mathbf{k}+b_{z}\sigma ^{z}\right] ~$I$_{\tau }\otimes $I$_{\rho }$, and
commutes with the Hamiltonian, we can replace it by its eigenvalues, $\Sigma
_{\pm }$. We find%
\begin{eqnarray}
&&\pm \Sigma _{\pm }=  \notag \\
&&\pm \left( \frac{1}{2}(\Delta _{1}^{2}+\Delta _{2}^{2})+\Delta
_{s}^{2}\mp \sqrt{\frac{1}{4}(\Delta _{1}^{2}-\Delta
_{2}^{2})^{2}+(\Delta _{1}^{2}+\Delta _{2}^{2})\Delta
_{s}^{2}+2\Delta _{s}^{2}\Delta _{1}\Delta _{2}\cos
(k_{z}d)}\right) ^{\frac{1}{2}},  \label{Sigma_}
\end{eqnarray}%
and
\begin{equation}
H_{\pm \pm }=\hbar v_{F}({\hat{z}}\times \mathbf{\sigma })\cdot \mathbf{k}%
+b_{z}\sigma ^{z}\pm \Sigma _{\pm }\sigma ^{z}.  \label{H_}
\end{equation}%
The first $\pm $ and second $\pm $ subscripts in $H_{\pm \pm }$ in (\ref{H_}%
) denote the $\pm $ in front of $\Sigma _{\pm }$,~and the subscript of $%
\Sigma _{\pm }$, respectively.

The energy eigenvalues are
\begin{equation}
E=\pm \sqrt{\hbar ^{2}v_{F}^{2}\mathbf{k}_{\perp }^{2}+(b_{z}\pm \Sigma
_{\pm })^{2}},
\end{equation}%
where $\pm \Sigma _{\pm }$ is in (\ref{Sigma_}).

The Weyl node happens at, for example, when $b_{z}-\Sigma _{-}=0$.$~$For
simplicity, we now denote $b_{z}$ as $b$. The locations are
\begin{eqnarray}
k_{z} &=&\frac{2\pi }{d}\pm k_{0}, \\
k_{0} &=&\frac{2}{d}\arccos (\mp \frac{1}{2\sqrt{\Delta _{1}\Delta _{2}}%
\Delta _{s}}[b^{4}-2(\Delta _{1}^{2}+\Delta _{2}^{2}+2\Delta
_{s}^{2})b^{2}+4(\Delta _{1}\Delta _{2}+\Delta _{s}^{2})^{2}]^{\frac{1}{2}}).
\end{eqnarray}%
This state is very similar to a wave-packet of electrons whose $k_{z}~$%
centers around the particular value $\frac{2\pi }{d}\pm k_{0}.~$In this
situation, the Weyl nodes are centered around $(k_{x},k_{y},k_{z})=(0,0,%
\frac{2\pi }{d}\pm k_{0})$.$~$So this means that the material only allows
the transmission of the state with prescribed range of momentum. The limit $%
\Delta _{2}/\Delta _{1}\rightarrow 1$ reduces to previous discussion, \cite%
{BuBa, BuHoBa} in which case both $\Delta _{1},\Delta _{2}~$equal to $\Delta
_{d}$.

The expansion around the band touching point $b_{z}-\Sigma _{-}=0$ gives
\begin{eqnarray}
H_{--} &=&\hbar v_{F}({\hat{z}}\times \mathbf{\sigma })\cdot \mathbf{k}\pm
\hbar v_{{\small 0}}k_{z}\sigma ^{z} \\
&=&\hbar v_{F}(\sigma ^{x}k_{y}-\sigma ^{y}k_{x})\pm \hbar v_{{\small 0}%
}\sigma ^{z}k_{z}  \label{H_Weyl_}
\end{eqnarray}%
where we redefined $k_{z}-(\frac{2\pi }{d}-k_{0})\rightarrow k_{z},$ and
where%
\begin{equation}
\hbar v_{{\small 0}}=\frac{\Delta _{s}^{2}\Delta _{1}\Delta _{2}\frac{d}{2b}%
\sin (k_{0}d)}{\Delta _{s}^{2}-b^{2}+\frac{1}{2}(\Delta _{1}^{2}+\Delta
_{2}^{2})}.
\end{equation}

Near the Weyl points, the dispersion is
\begin{equation}
E(\mathbf{k})=\pm \hbar \sqrt{v_{F}^{2}(k_{x}^{2}+k_{y}^{2}\mathbf{)}+v_{%
{\small 0}}^{2}k_{z}^{2}}.
\end{equation}%
This is a Weyl fermion in three spatial dimensions, as described in (\ref%
{H_Weyl_}). This Weyl fermion is half of the components of a Dirac fermion.
These Weyl fermions come in pairs, around $k_{z}~$=$\frac{2\pi }{d}\pm k_{0}$%
. The two Weyl fermions are separated in the $\mathbf{k}$ space and have
opposite helicities. The heterostructure of topological insulators and
normal insulators thus become Weyl semimetals.

If $\Delta _{2}/\Delta _{1}\rightarrow 1,~$the above reduces to
\begin{equation}
\Sigma _{\pm }=\sqrt{\Delta _{1}^{2}+\Delta _{s}^{2}\mp 2\Delta
_{s}\Delta _{1}\cos (\frac{1}{2}k_{z}d)},
\end{equation}%
The branches with different signs can be understood as a phase shift in $%
\frac{1}{2}k_{z}d\rightarrow \frac{1}{2}k_{z}d+\pi .~$The period when $%
\Delta _{2}\neq \Delta _{1}$ is $\frac{2\pi }{d},$ and when $\Delta
_{2}/\Delta _{1}\rightarrow 1~$is enhanced to$~\frac{4\pi }{d}=\frac{2\pi }{%
d/2}.~$When taking the $\Delta _{2}/\Delta _{1}\rightarrow 1~$limit, $\Delta
_{1}=\Delta _{2}=\Delta _{d},~$%
\begin{eqnarray}
\Delta _{1}^{2}+\Delta _{s}^{2}-b^{2} &=&2\Delta _{s}\Delta _{1}\cos (\frac{1%
}{2}k_{0}d), \\
\hbar v_{0} &=&\Delta _{s}\Delta _{1}\frac{d}{2b}\sin (\frac{1}{2}k_{0}d)
\end{eqnarray}%
which reduces to previous discussion \cite{BuBa, BuHoBa} in the $\Delta
_{2}/\Delta _{1}\rightarrow 1~$limit.

When we tune the parameters $b$ or the tunnelings,~the band touching
occurred and the transition\cite{BuBa} between insulators and Weyl
semimetals is closely related to the topology change of the band structure
on the parameter space.

There is another limit $\Delta _{2}/\Delta _{1}\rightarrow 0~$limit, in
which the tunneling across the NI$^{\prime }$ material is taken to zero, so
this is the limit when there is TI/NI/TI heterojunction.

The doping with magnetic impurities can be experimentally performed in, for
example,\cite{Chen et al} by for example \textrm{Mn} doped \textrm{Bi}$_{2}%
\mathrm{Se}_{3}.$ These heterostructures can be experimentally made by using
topological insulator thin films, which have been demonstrated in
experiments.\cite{Zhang G et al}

These models of Weyl semimetals have bulk gapless modes at particular
momentum vectors, near the above Weyl nodes, and these modes with the
particular momentum vectors near the Weyl nodes conduct electric current as
well as thermal current. These materials have electrical conductivity and
thermal conductivity which can be measured and useful.

\section{Heterostructure of TI/SC/TI/SC$^{\prime }$}


\label{sec_ heterostructure TI/SC/TI/SC'}

One can also replace the normal insulator materials discussed in the last
section with superconductors (SC), and consider TI/SC/TI/SC$^{\prime }$
heterostructures. One can also consider the model by changing the NI to
superconductor (SC), and construct the model corresponding to TI/SC/TI/SC$%
^{\prime }$ heterostructure. The superconductors can introduce
couplings between opposite spins within the same surfaces (for
both upper surface and lower
surface), $|\Delta |e^{i\varphi }c_{\boldsymbol{k}\uparrow }^{\dag }c_{-%
\boldsymbol{k}\downarrow }^{\dag }+$h.c.. The parameters for SC and SC$%
^{\prime }$ are slightly different.

Now the Hamiltonian is further enlarged by a $\kappa $-space. The $su(2)$ $%
\kappa $-space is due to particle-hole symmetry. The size or
dimension of the state vector is doubled due to particle-hole
symmetry. The periodic TI/SC model has been considered.\cite{MeBa}
The surface Hamiltonian of each surface is further perturbed by a
superconducting pairing term.

We again transform the Hamiltonian to momentum space via (\ref{transform_}).
We then make the transformation $\tau ^{\pm }\rightarrow \tau ^{\pm }\sigma
^{z},\sigma ^{\pm }\rightarrow \sigma ^{\pm }\tau ^{z}$. The size of the
Hamiltonian is $16\times 16$. For TI/SC/TI/SC$^{\prime }$ heterostructure,
the model Hamiltonian in momentum space is
\begin{equation}
H=\sum_{\boldsymbol{k};{\hat{\imath}},{\hat{\jmath}}=\pm ,\pm }c_{%
\boldsymbol{k}{\hat{\imath}}}^{\dag }H_{{\hat{\imath}\hat{\jmath}}}c_{%
\boldsymbol{k}{\hat{\jmath}}}+\sum_{\boldsymbol{k};{\hat{\imath}}={\hat{%
\jmath}}=\pm ,\pm }(|\Delta |e^{i\varphi }c_{\boldsymbol{k}{\hat{\imath}}%
\uparrow }^{\dag }c_{-\boldsymbol{k}{\hat{\jmath}}\downarrow }^{\dag }+\text{%
h.c.})
\end{equation}%
where the second term is superconducting pairing term with $\Delta =|\Delta
|e^{i\varphi }.~H_{{\hat{\imath}\hat{\jmath}}}$ is the Hamiltonian without
adding the superconducting term, and ${\hat{\imath},\hat{\jmath}}$ denote
four blocks corresponding to $\pm ,\pm $ in (\ref{H_}).

These four eigenvalues of $H_{{\hat{\imath}\hat{\jmath}}}$ correspond to
sectors of $\tau ^{z},\rho ^{z}=\pm 1,\pm 1,$%
\begin{equation}
H_{\pm \pm }=\hbar v_{F}({\hat{z}}\times \mathbf{\sigma })\cdot \mathbf{k}%
+(b_{z}\pm \Sigma _{\pm })\sigma ^{z}
\end{equation}%
with the eigenvalues
\begin{eqnarray}
&&(b_{z}\pm \Sigma _{\pm })  \notag \\
&=&b\pm \left( \frac{1}{2}(\Delta _{1}^{2}+\Delta _{2}^{2})+\Delta
_{s}^{2}\mp \sqrt{\frac{1}{4}(\Delta _{1}-\Delta _{2})^{2}+(\Delta
_{1}^{2}+\Delta _{2}^{2})\Delta _{s}^{2}+2\Delta _{s}^{2}\Delta
_{1}\Delta
_{2}\cos (k_{z}d)}\right) ^{\frac{1}{2}}.  \notag \\
&&  \label{b_sigma}
\end{eqnarray}

When adding the superconducting term, the $4\times 4$ Hamiltonians are
\begin{eqnarray}
H_{i=\pm \pm }^{\Delta } &=&\frac{1}{2}\sum_{\boldsymbol{k}}\psi
_{k,i}^{\dag }(\hbar v_{F}({\hat{z}}\times \mathbf{\sigma })\cdot \mathbf{k}~%
\text{I}_{\kappa }+\sigma ^{z}[(b_{z}\pm \Sigma _{\pm })\text{I}_{\kappa }+%
\frac{1}{2}(|\Delta |e^{i\varphi }\kappa ^{+}+|\Delta |e^{-i\varphi }\kappa
^{-})])\psi _{k,i}  \notag \\
&&  \label{H_t_sc}
\end{eqnarray}%
where the basis is $\psi _{\boldsymbol{k},i}=(c_{\boldsymbol{k}i\uparrow
},c_{\boldsymbol{k}i\downarrow },c_{-\boldsymbol{k}i\downarrow }^{\dag },c_{-%
\boldsymbol{k}i\uparrow }^{\dag }),~i=(\pm ,\pm ).~$The superscript $\Delta $
in $H_{i=\pm \pm }^{\Delta }$ denotes including the superconducting term.
The first $\pm $ and second $\pm $ subscripts in $H_{i=\pm \pm }^{\Delta }$
in (\ref{H_t_sc}) denote $\tau ^{z},\rho ^{z}=\pm 1,\pm 1$. There are four
blocks corresponding to $(\pm ,\pm )$.~The$~b_{z}\sigma ^{z}$ term is a
deformation of the quadratic term, and is due to the doping of magnetic
impurities in the TI. $\kappa ^{x},\kappa ^{y},\kappa ^{z}~$are Pauli
matrices, and $\kappa ^{+}=\kappa ^{x}+i\kappa ^{y},\kappa ^{-}=\kappa
^{x}-i\kappa ^{y}$. The superconducting term was from a quartic term, and
after a mean field treatment it becomes a quadratic term in the Hamiltonian.

For example, for $\tau ^{z},\rho ^{z}=-1,-1,$%
\begin{eqnarray}
H_{--}^{\Delta } &=&\frac{1}{2}\sum_{\boldsymbol{k}}\psi _{\boldsymbol{k,}%
{\normalsize --}}^{\dag }(\hbar v_{F}({\hat{z}}\times \mathbf{\sigma })\cdot
\mathbf{k~}\text{I}_{\kappa }+\sigma ^{z}[(b_{z}-\Sigma _{-})\text{I}%
_{\kappa }+\frac{1}{2}(|\Delta |e^{i\varphi }\kappa ^{+}+|\Delta
|e^{-i\varphi }\kappa ^{-})])\psi _{\boldsymbol{k}{\normalsize ,--}}.  \notag
\\
&&
\end{eqnarray}

In the matrix form in $\kappa $-space,%
\begin{equation}
H_{i=\pm \pm }^{\Delta }(\mathbf{k})=\left[
\begin{array}{cc}
\left[ \hbar v_{F}({\hat{z}}\times \mathbf{\sigma })\cdot \mathbf{k}%
+(b_{z}\pm \Sigma _{\pm })\sigma ^{z}\right] ~ & |\Delta |e^{i\varphi
}\sigma ^{z} \\
|\Delta |e^{-i\varphi }\sigma ^{z} & \left[ \hbar v_{F}({\hat{z}}\times
\mathbf{\sigma })\cdot \mathbf{k}+(b_{z}\pm \Sigma _{\pm })\sigma ^{z}\right]%
\end{array}%
\right] .  \label{H_kappa}
\end{equation}

The Hamiltonian in (\ref{H_kappa}) can be diagonalized in $\kappa
$-space. When it is diagonalized in $\kappa $-space, this is given
by a transformation
\begin{equation}
H=\frac{1}{2}\sum_{\boldsymbol{k};i,j=(\pm ,\pm )}\tilde{\psi}_{i,%
\boldsymbol{k}}^{\dag }H_{ij}^{\Delta }\tilde{\psi}_{j,\boldsymbol{k}},
\end{equation}%
with the basis of state vector%
\begin{equation*}
(\tilde{\psi}_{+,\boldsymbol{k}},\tilde{\psi}_{+,(-\boldsymbol{k})}^{\dagger
},\tilde{\psi}_{-,\boldsymbol{k}},\tilde{\psi}_{-,(-\boldsymbol{k}%
)}^{\dagger }),
\end{equation*}%
where\cite{MeBa}%
\begin{eqnarray}
\tilde{\psi}_{+,\boldsymbol{k}} &=&\frac{1}{\sqrt{2}}e^{-\frac{1}{2}i\varphi
}c_{\boldsymbol{k}i\uparrow }+\frac{1}{\sqrt{2}}e^{\frac{1}{2}i\varphi }c_{-%
\boldsymbol{k}i\downarrow }^{\dag }, \\
\tilde{\psi}_{-,\boldsymbol{k}} &=&-\frac{i}{\sqrt{2}}e^{-\frac{1}{2}%
i\varphi }c_{\boldsymbol{k}i\uparrow }+\frac{i}{\sqrt{2}}e^{\frac{1}{2}%
i\varphi }c_{-\boldsymbol{k}i\downarrow }^{\dag },
\end{eqnarray}

It can again be diagonalized in $\kappa $-space. For $\tau ^{z},\rho
^{z}=-1,-1,$ for~ $\kappa ^{z}=\pm 1$,
\begin{equation}
H_{--}^{\Delta \pm }=\hbar v_{F}({\hat{z}}\times \mathbf{\sigma })\cdot
\mathbf{k}+\sigma ^{z}(b_{z}-\Sigma _{-}\pm |\Delta |),  \label{H_del}
\end{equation}%
where the superscript $\pm $ in $H_{--}^{\Delta \pm }$ denotes $\kappa
^{z}=\pm 1$,$~$and $\pm |\Delta |$ in (\ref{H_del}) correspond to $\kappa
^{z}=\pm 1.$

Similarly for the four sectors $\tau ^{z},\rho ^{z}=\pm 1,\pm 1$,
\begin{equation}
H_{\pm \pm }^{\Delta \pm }=\hbar v_{F}({\hat{z}}\times \mathbf{\sigma }%
)\cdot \mathbf{k}+\sigma ^{z}(b_{z}\pm \Sigma _{\pm }\pm |\Delta |).
\end{equation}%
$|\Delta |$ effectively shifts $b_{z}\pm \Xi _{\pm }$. The eigenvalue is
\begin{equation}
E(\mathbf{k})=\pm \sqrt{\hbar ^{2}v_{F}^{2}\mathbf{k}_{\perp }^{2}+(b_{z}\pm
\Sigma _{\pm }\pm |\Delta |)^{2}}.
\end{equation}%
where $b_{z}\pm \Sigma _{\pm }$ are in (\ref{b_sigma}).

The Bogoliubov-Weyl nodes near the band touching points are located at for
example $b_{z}-\Sigma _{\pm }\pm |\Delta |=0.~$For $b_{z}-\Sigma _{-}\pm
|\Delta |=0,$ they are%
\begin{eqnarray}
k_{z} &=&\frac{2\pi }{d}\pm k_{0}^{\Delta }, \\
k_{0}^{\Delta } &=&\frac{2}{d}\arccos (\mp \frac{1}{2\sqrt{\Delta _{1}\Delta
_{2}}\Delta _{s}}[(b\pm |\Delta |)^{4}-2(\Delta _{1}^{2}+\Delta
_{2}^{2}+2\Delta _{s}^{2})(b\pm |\Delta |)^{2}+4(\Delta _{1}\Delta
_{2}+\Delta _{s}^{2})^{2}]^{\frac{1}{2}})  \notag \\
&&
\end{eqnarray}%
and the Fermi velocity is%
\begin{equation}
\hbar v_{0}=\frac{\Delta _{s}^{2}\Delta _{1}\Delta _{2}\frac{d}{2b}\sin
(k_{0}d)}{\Delta _{s}^{2}-(b\pm |\Delta |)^{2}+\frac{1}{2}(\Delta
_{1}^{2}+\Delta _{2}^{2})}.
\end{equation}%
We have thus found the energy eigenvalues, the locations of Weyl
nodes, and the Fermi velocity near the node.

Interestingly, Weyl superconducting phases can also be realized 
by triplet pairing phases.\cite{LiWu, Sau Tewari}


\vspace{1pt}

\section{Other relevant materials and Discussion}

\label{sec_ other relevant materials}


The $su(2)$ geometric phase with two Chern numbers may occur also in other
condensed matter systems or atomic and molecular systems. There are related
discussions in other possible situations.\cite{Wu,Murakami etal, Chang etal}
The Semiconductor model includes at least light hole (LH) and heavy hole
(HH) bands. If both LH and HH are doubly degenerate, they can also exhibit $%
su(2)$ geometric phases with two Chern numbers, in the $d$-space. There are
related discussions\cite{Murakami etal, Chang etal} to these aspects. It is
possible to exhibit general value of Chern numbers, by for example
multilayer heterostructures of particular semiconductors.

The cold atomic system with cold atom of a larger total angular momentum of $%
F=3/2$, can also exhibit $su(2)$ geometric phase in the parameter space of
paring condensates, for example.\cite{Wu} It may also be relevant to
Interpenetrating Lattices of two optical lattices of cold atom systems.

It may be interesting to see whether these fibrations can be realized
concretely in these experimental settings.

%

\acknowledgements

This work was supported in part by NSF grant DMS-1159412, and NSF
grant DMS-0804454, and also in part by the Fundamental Laws
Initiative of the Center for the Fundamental Laws of Nature,
Harvard University. We thank S. T. Chui for correspondences.

\vspace{1pt}

{\normalsize \ }

\end{document}